\documentclass[12pt, prd, showpacs]{revtex4}
\usepackage{amsmath}
\begin{document}

\title{Quasiblack holes with pressure: General exact results}
\author{Jos\'{e} P. S. Lemos}
\affiliation{Centro Multidisciplinar de Astrof\'{\i}sica, CENTRA,
Departamento de F\'{\i}sica, Instituto Superior T\'ecnico - 
IST, Universidade T\'{e}cnica de Lisboa
- UTL, Av. Rovisco Pais 1, 1049-001 Lisboa, Portugal\,\,
\\\&\\
Institute of Theoretical Physics - ITP, Freie Universit\"at Berlin,
Arnimallee 14 D-14195 Berlin, Germany.}
\email{joselemos@ist.utl.pt}
\author{Oleg B. Zaslavskii}
\affiliation{Astronomical Institute of Kharkov V.N. Karazin National
University, 35
Sumskaya St., Kharkov, 61022, Ukraine}
\email{ozaslav@kharkov.ua}

\begin{abstract}
A quasiblack hole is an object in which its boundary is situated at a
surface called the quasihorizon, defined by its own gravitational radius. We
elucidate under which conditions a quasiblack hole can form under the
presence of matter with nonzero pressure. It is supposed that in the outer
region an extremal quasihorizon forms, whereas inside, the quasihorizon can
be either nonextremal or extremal. It is shown that in both cases,
nonextremal or extremal inside, a well-defined quasiblack hole always
admits a continuous pressure at its own quasihorizon. Both the nonextremal
and extremal cases inside can be divided into two situations, one in which
there is no electromagnetic field, and the other in which there is an
electromagnetic field. The situation with no electromagnetic field requires
a negative matter pressure (tension) on the boundary. On the other hand, the
situation with an electromagnetic field demands zero matter pressure on the
boundary. So in this situation an electrified quasiblack hole can be
obtained by the gradual compactification of a relativistic star with the
usual zero pressure boundary condition. For the nonextremal case inside the
density necessarily acquires a jump on the boundary, a fact with no harmful
consequences whatsoever, whereas for the extremal case the density is
continuous at the boundary. For the extremal case inside we also state and
prove the proposition that such a quasiblack hole cannot be made from
phantom matter at the quasihorizon. The regularity condition for the
extremal case, but not for the nonextremal one, can be obtained from the
known regularity condition for usual black holes.
\end{abstract}

\maketitle



\newpage

\section{Introduction}

In recent years, the taxonomy of relativistic objects has increased to
include the so-called quasiblack holes. The general definition and
description of the general properties of these objects can be found in
\cite
{qbh}. Here, we recall that a quasiblack hole is, roughly speaking, an
object on the verge of forming a horizon but without collapsing, so the
system remains static even when the boundary approaches its own
gravitational radius surface, or the quasihorizon, as nearly as one likes.
It turns out that nonextremal quasiblack holes are connected with the
appearance of diverging surface stresses when the boundary approaches the
quasihorizon, so only extremal quasiblack holes are free from infinite
surface stresses.

The significance of quasiblack holes is twofold. First, it is a useful
methodical tool for better understanding the general features of black holes
such like the relation to black hole mimickers \cite{wh}, the mass formula
\cite{mass1,mass2} and entropy \cite{entr,extrentr}. In doing so, one should
not bother about the physical realization of such construction and even
admit infinite surface stresses to obtain finite final formulas for physical
quantities (see \cite{mass1}). Second, quasiblack holes can be of interest
by themselves, as real physical objects. There are several examples of
objects that exhibit quasiblack hole behavior. Simple systems, which can
be treated analytically, like Bonnor stars, made of Majumdar-Papapetrou
matter, i.e., extremal dust where the density of matter is equal to that of
the charge so that the matter pressure is zero, matched to an extreme
Reissner-Nordstr\"om vacuum, admit quasiblack holes \cite
{bon,lzanchin1,lzanchin2}. Continuous Majumdar-Papapetrou systems made
purely from extremal dust also admit quasiblack holes \cite{extdust}. More
complex structures like self-gravitating Yang-Mills--Higgs magnetic
monopoles also possess quasiblack holes, as found previously in \cite
{lw1,lw2}.

In \cite{g} exact relativistic charged sphere solutions with pressure were
found. Drawing upon this work on exact solutions \cite{g} and upon previous
work on charged systems with pressure \cite{jz0}, it was shown in \cite{jz}
that there are electrically charged quasiblack holes with pressure which are
obtained as limiting cases of the relativistic charged spheres of \cite{g},
namely, these quasiblack holes can be thought of as being formed when a star
made of charged matter with pressure is sufficiently compressed. In the
study \cite{jz}, the corresponding models have the attractive feature that in
some range of parameters the speed of sound is real and less than that of
light. In \cite{felice1,felice2}, numerical work was performed on a different
but similar type of relativistic charged spheres which degenerates into
quasiblack holes with pressure when the spheres are sufficiently compact.
The study of pressure charged systems not only extends the class of
electrically charged quasiblack holes but also brings an important feature
connected with the issue of stability to those systems. The point is that
quasiblack holes made purely from extremal dust, are unstable with respect
to a dynamic perturbation having kinetic energy. With the presence of
pressure, there is the possibility of finding stable configurations. Indeed,
in \cite{felice2} it was found that there were instances in which the
systems are stable against radial perturbations, and this might indicate
that the quasiblack holes found in \cite{jz} are also stable. The
self-gravitating Yang-Mills--Higgs magnetic monopole quasiblack holes
studied in \cite{lw1,lw2} can be considered as quasiblack holes with
pressure since an intrinsic inbuilt effective pressure is present in the
Yang-Mills--Higgs equations, and thus, might also be stable systems.

Following our previous works
\cite{qbh,wh,mass1,mass2,entr,extrentr}, we want to put forward a
general model-independent approach and find the conditions under which
quasiblack holes, extremal to the outside, with pressure are
possible. We work with quasiblack holes that are extremal from the
outside because only these are regular and free from infinite surface
stresses, nonextremal quasiblack holes having diverging surface
stresses \cite{mass1}. The study is quite general, in the sense that
the outside extremality condition can be of any type, it can be due to
a specific mass to charge relation, or to a specific mass to
cosmological constant relation, to name two cases among others. If,
for instance, the external region is described by the
Reissner-Nordstr\"{o}m metric, its charge $q$ is equal to mass $m$,
$q=m$. On the other hand, from inside we allow that the quasihorizon
can be either nonextremal or extremal. Nonextremal quasihorizons from
the inside with matter pressure were found in \cite{jz}.  Extremal
quasihorizons with pressure for self-gravitating magnetic monopoles
were studied in \cite{lw1,lw2}. Our analysis includes all these
systems and extends to pressure systems the pressureless cases treated
in [1]. Moreover, we treat the cases in which from the outside the
quasihorizon is always extremal whereas from the inside the
quasihorizon can be either nonextremal or extremal.

This paper is organized as follows: In Sec.~\ref{bf}, we write the basic
formulas for a generic spherically symmetric system and for the system
when it is in a state of transition to a quasiblack hole. In
Sec.~\ref{qbhp}, we make a deep analysis of the conditions on the
radial pressure the quasihorizon of a quasiblack hole must obey in the
cases where there is an nonextremal quasihorizon from the inside and an
extremal quasihorizon from the inside. We also study the conditions on
the energy density and make some comments related to the null energy
condition. In Sec.~\ref{conc}, we conclude.

\section{Basic formulas and limiting transition}

\label{bf}

\subsection{Basic formulas}

Consider a metric $g_{\mu \nu }$ with line element $ds^{2}=g_{\mu \nu
}\,dx^{\mu }\,dx^{\nu }$ for a spherically-symmetric spacetime containing
matter, i.e.,
\begin{equation}
ds^{2}=-U(r)\,dt^{2}+V(r)^{-1}dr^{2}+r^{2}\left( d\theta ^{2}+\sin
^{2}\theta d\phi ^{2}\right) \text{.}
\end{equation}
The stress-energy tensor of the matter has the form
\begin{equation}
{T_{\mu }}^{\nu }=\mathrm{diag}(-\rho ,p_{r},p_{\perp },p_{\perp })\,,
\end{equation}
where $\rho $, $p_{r}$ and $p_{\perp }$ are the energy density, the radial
pressure, and the tangential pressure, respectively. The Einstein equations
are $G_{\mu \nu }=8\pi \,T_{\mu \nu }$, where $G_{\mu \nu }$ is the Einstein
tensor and $G=1$, $c=1$ here. The two equations of interest are the $tt$ and
$rr$ components. If we put
\begin{equation}
U(r)=V(r)\,\exp (2\psi (r))\text{,}  \label{u}
\end{equation}
then it follows from the Einstein equations that
\begin{equation}
2\psi (r)=\int^{r}d\bar{r}\,\frac{\sigma (\bar{r})}{V(\bar{r})}\text{,}
\end{equation}
where we have defined the quantity $\sigma (r)$ as
\begin{equation}
\sigma (r)=8\pi r\bigl(p_{r}(r)+\rho (r)\bigr)\,.  \label{sigmag}
\end{equation}
And if we put
\begin{equation}
V(r)= 1-\frac{2m(r)}{r}\text{,}  
\label{v}
\end{equation}
then it follows that
\begin{equation}
m(r)=4\pi \int_{0}^{r}d\bar{r}\,\bar{r}^{2}\rho (\bar{r})\text{.}  \label{m}
\end{equation}
Here, we assume that the center $r=0$ is a regular one, and there is no
horizon a priori.

Let us consider a compact body situated in the inside region such that $
r\leq r_{0}$. The radius $r=r_{0}$ defines the boundary which divides the
inside region from the outside one. We do not specify the metric outside,
for $r> r_{0}$. In particular, it can be the Reissner-Nordstr\"{o}m metric.
In what follows we will use subscripts \textquotedblleft in" and
\textquotedblleft out" to distinguish quantities in each of
the two regions. To
match the two metrics, i.e., the first quadratic forms, at the boundary $
r=r_{0}$, we need the condition
\begin{equation}
U_{\mathrm{in}}(r_{0})=U_{\mathrm{out}}(r_{0})\,\text{.}  \label{a0}
\end{equation}
We assume that there is no massive shell on the boundary, which entails the
continuity of the metric potential $V$,
\begin{equation}
V_{\mathrm{in}}(r_{0})=V_{\mathrm{out}}(r_{0})\text{.}
\end{equation}
In addition, without essential loss of generality, we deal with
metrics for which $U_{\mathrm{out}}(r)=V_{\mathrm{out}}(r)$, since
this simplifies the formulas. In particular, the
Reissner-Nordstr\"{o}m metric belongs to this class, in which case
$U_{\mathrm{out}}(r)=1-\frac{2m(r)}{r}$, with $ m(r)=m-q^2/2r$, and in
the extremal case $m=q$, we are interested in one that has $
m(r)=m-m^2/2r$, so that $U_{\mathrm{out}}(r)=(1-m/r)^2$. Then, after
simple manipulations, we obtain that
\begin{equation}
U_{\mathrm{in}}(r)=V_{\mathrm{in}}(r)\exp \left(2\psi (r_{0},r)\right)\,,
\label{uin0}
\end{equation}
with
\begin{equation}
2\psi (r_{0},r)= \int_{r_{0}}^{r}d\bar{r} \frac{\sigma(\bar{r})}{V(\bar{r})}
\text{.}  
\label{uin}
\end{equation}
We do not specify further properties beforehand, in particular, the
presence of transverse surface stresses is allowed.

\subsection{Limiting transition}

Now we make the next assumption, namely, there is a limiting transition in
the course of which a horizon almost forms. From (\ref{a0}) one can then
write
\begin{equation}
U_{\mathrm{in}}(r_{0})=U_{\mathrm{out}}(r_{0})\equiv U(r_{0})=\varepsilon
\,,
\label{uin1}
\end{equation}
where $\varepsilon $ is any number that can be made as small as one wants,
$
\varepsilon <<1$. Since we are interested in the limit $\varepsilon
\rightarrow 0$, this means that the quantity $U(r_{0})=\varepsilon $ becomes
a small parameter and the areal radius $r_{0}$ approaches the radius of a
would-be-horizon $r_{+}$. We want to examine whether and under which
condition a quasiblack hole can appear. By itself, the proximity of $r_{0}$
to $r_{+}$ is insufficient. It is also required that in the whole inner
region $r\leq r_{0}$ the lapse function $U_{\mathrm{in}}(r)\rightarrow 0$ in
such a way that
\begin{equation}
U_{\mathrm{in}}(r)=\varepsilon f(r)\,,  \label{f}
\end{equation}
where $f(r)$ is some bounded function. Furthermore, $f(r_{+})\neq 0$. The
latter condition is needed to distinguish a quasiblack hole from a true
black hole. More exactly, this function must obey the condition
$f(r_{+})=1$, 
as is seen from (\ref{uin1}) and (\ref{f}). Formally, we can also admit
a nonmonotonic $f(r)$ which inside, in some subregion, is of the order $
\varepsilon ^{-\gamma }$ with $0<\gamma<1$. Then $U\rightarrow 0$ everywhere
inside. However, for the most physically interesting cases of quasiblack
holes, $U(r)$ is a monotonically decreasing function of $r$, see Appendix B
of \cite{qbh}.

From an outside perspective, the supposed quasihorizon can be, in principle,
nonextremal or extremal. From a physical viewpoint, the latter case is more
important since it is the extremal quasiblack hole case which is indeed
regular \cite{qbh}, whereas the nonextremal quasiblack hole case leads to
infinite surface stresses \cite{mass1}. Thus, we assume that to the outside
the quasiblack hole is extremal. The study is valid for any extremal type of
outside horizon. In the situation there is an extremal electrically charged
horizon then the charge equals the mass, $q=m$.

Now, even being extremal to the outside, the quasiblack hole can have
a horizon which, from the inside, is either nonextremal or
extremal. Indeed, an extremal horizon for outside observers implies
that the metric potential $ V(r)$ has in the limit a double root when
considered from outside. However, as shown in a concrete example in
\cite{jz}, from inside the horizon can be either nonextremal or
extremal. Therefore, we will consider the two cases separately, i.e.,
we will consider first quasiblack holes with a nonextremal horizon
from the inside, and second quasiblack holes with an extremal horizon
from the inside. Both are extremal quasiblack holes from the outside.

\section{Quasiblack holes with pressure}

\label{qbhp}

\subsection{Quasiblack holes with pressure, nonextremal from the inside}

\subsubsection{General considerations}

In the nonextremal from the inside case, near the gravitational radius of
the configuration, the asymptotic form of the metric potential $V$ inside
should be
\begin{equation}
V_{\mathrm{in}}=\varepsilon +k(r_{0}-r)+...\,,  \label{as}
\end{equation}
with $\varepsilon<<1$, $k>0$, $k$ being some quantity with units of inverse
length. See \cite{jz} for concrete examples of this case of quasiblack holes
with pressure, nonextremal from the inside. We want to elucidate the
conditions on the parameters of the system, when the quantity $U$ is
uniformly bounded everywhere inside, i.e., is of the form (\ref{f}). We
analyze first the behavior of the functions in the bulk of the matter $
r<r_{0}$, and second at the boundary $r_{0}$, in both cases we assume that
the quasiblack hole is being formed, $r_{0}\rightarrow r_{+}$.

\textit{Region in the bulk of the matter, $r<r_{0}$.} To this end, let us
rewrite Eqs.~(\ref{uin0})-(\ref{uin}) in the form
\begin{equation}
U_{\mathrm{in}}=V_{\mathrm{in}}P_{1}P_{2}\,.
\end{equation}
Here
\begin{equation}
P_{1}=\exp (2\psi _{1})\;\text{, }\quad 2\psi _{1}= \int_{r_{0}}^{r}d\bar{r}
\,\frac{\sigma_0}{V_{\mathrm{in}}({\bar{r}})}\,,  \label{xi->2psi1}
\end{equation}
\begin{equation}
P_{2}=\exp (2\psi _{2})\;\text{, }\quad 2\psi _{2}= \int_{r_{0}}^{r}d\bar{r}
\,\frac{{\sigma(\bar r)}- \sigma_0}{V_{\mathrm{in}}({\bar{r}})}\,\text{,}
\label{u2}
\end{equation}
where $\sigma(r)$, defined in Eq.~(\ref{sigmag}), is a quantity with units
of surface density (i.e., inverse length) and $\sigma_0 \equiv \sigma (r_0)$
in an obvious notation. It is also useful to define $\sigma_+ \equiv \sigma
(r_+)$, i.e.,
\begin{equation}
\sigma_+ =8\pi r_{+}\bigl(p_{r}(r_{+})+\rho (r_{+})\bigr)\text{.}  \label{a}
\end{equation}
Taking into account (\ref{as}), we see that $\lim_{\varepsilon \rightarrow
0}P_{2}$ is a well-defined nonzero quantity that remains everywhere bounded
including the boundary $r=r_{0}=r_{+}$. Let us focus attention on $P_{1}$.
Then, one can write $\psi _{1}$ in the form
\begin{equation}
2\psi _{1}(r)=\frac{\sigma _{+}}{k}\bigl(\ln \varepsilon +2\psi _{1\epsilon
}(r)\bigr)+2\psi _{11}(r)  \label{q}
\end{equation}
where $2\psi _{1\epsilon }(r)=-\ln \left(\varepsilon +k(r_{0}-r)\right)$. It
follows from (\ref{a}) and the asymptotic behavior (\ref{as}) that in the
limit when $r_{0}\rightarrow r_{+}$ (that entails $\varepsilon \rightarrow
0$) 
the quantity $2\psi _{11}(r)$ is finite everywhere inside, including the
limit $\varepsilon =0$, $2\psi _{11}(r_{0})=0$. Making the rescaling of time
according to \thinspace $T=t\left(\frac{\varepsilon}{k}\right) ^{\frac{
\sigma_{+}}{2k}}$, we obtain inside the metric
\begin{equation}
ds^{2}=-\frac{V(r)}{\left(\varepsilon +k(r_{0}-r)\right)^{\sigma _{+}/k}}
\,g(r)\,dT^{2}+ \frac{dr^{2}}{V(r)}+r^{2}\left( d\theta ^{2}+\sin ^{2}\theta
d\phi ^{2}\right)  \label{g}
\end{equation}
where $g(r)\equiv \exp (2\psi _{11}+2\psi _{2})$ is everywhere finite and
does not vanish.

Now, we want to impose that the metric (\ref{g}) be free of curvature
singularities by requiring that in an orthonormal frame the components of
the Riemann tensor be finite. There is only one such potentially divergent
term for the metric (\ref{g}). It is the component
\begin{equation}
{R_{0r}}^{0r}=-\frac{1}{4}\,{V^{\,\prime }\,(\ln
U)^{\,\prime }}-\frac{1}{4}
\,V\,\left( 2\,(\ln U)^{\,\prime \prime }+(\ln U)^{\,\prime \,2}\right) \,,
\label{e}
\end{equation}
where $U$ is the potential of $dT^{2}$ in (\ref{g}), and a $^{\,\prime }$
denotes a derivative with respect to the argument, in this case $r$. A
simple, but nontrivial, analysis shows that there are only two ways to
achieve finiteness in (\ref{e}). 
Indeed, using Eqs.~(\ref{v}), (\ref{uin0}), and (\ref{uin}) in (\ref{e})
one finds
\begin{equation}
{R_{0r}}^{0r}=K-Q\text{, }
\label{ge0}
\end{equation}
where 
\begin{equation}
K=-\left(\frac{m}{r^{2}}+4\pi
rp_{r}\right)^{\prime }\text{, }
\end{equation}
and 
\begin{equation}
Q= \frac{\sigma\left(\sigma +V^{\prime }\right)}{4V}. 
\label{ge}
\end{equation}
We want to exclude the presence of a shell, so we want to have the pressure
continuous. Then, $p_{r}^{\prime}$ is finite and so is the quantity $K$.
The potential divergences can be connected with the term $Q$ only. It
follows from Eq.~(\ref{as}) that in the limit under discussion
\begin{equation}
{Q}\approx \frac{\sigma _{+}\left( \sigma _{+}-k\right) }{4\,V}\,.
\label{ge14}
\end{equation}
There are thus two possibilities: either $\sigma _{+}=0$, which as we will
see yields the regular black hole, or $\sigma _{+}=k$, which yields the
quasiblack hole.

The first way is to put $\sigma _{+}=0$. Then, we get from (\ref{q}) $2\psi
_{1}(r)=2\psi _{11}(r)$, so that, as $P_{2}$ (see above), $P_{1}$ is finite.
Since $U=VP_{1}P_{2}$, it follows from (\ref{as}) that $U\sim r_{0}-r$, and
thus $U\sim V$. So, instead of a quasihorizon, in the limit of $\varepsilon
\rightarrow 0,$ $r_{0}\rightarrow r_{+}$, we obtain a regular event horizon
(see, e.g., \cite{visser}), of the type found in the Schwarzschild,
Reissner-Nordstr\"{o}m or generic regular black holes discussed in
\cite{dym}
(see also \cite{lake,bz}). In doing so, the metric coefficient
$U_{\mathrm{in
}}(r)$ does not have the form (\ref{f}). Thus, as we want to ensure the
existence of a quasiblack hole, we reject the choice $\sigma _{+}=0$.

The second way to achieve finiteness is to put $\sigma _{+}=k$. Then, in the
limit $\varepsilon \rightarrow 0$ one has $2\psi _{1}(r)=\ln \varepsilon +
\mathrm{finite\;terms}$, so that $P_{1}\sim \varepsilon $, and so also $
U\sim \varepsilon $, i.e., we obtain the metric function $U$ in the form 
(\ref{f}), the form appropriate for a quasiblack hole. Thus, we choose $
\sigma _{+}=k$. Using the expressions (\ref{v})-(\ref{m}), the equality $
r_{+}=2m(r_{+})$, and neglecting the difference between $r_{0}$ and $r_{+}$,
one obtains $V^{\,\prime }(r_{+})=-\left( 8\pi \rho ^{\mathrm{in}
}(r_{+})r_{+}- \frac{1}{r_{+}}\right) $. From Eq.~(\ref{as}) one has $
k=-V^{\,\prime }(r_{+})$, i.e., $k=8\pi \rho ^{\mathrm{in}}(r_{+})r_{+}-
\frac{1}{r_{+}}$. Then, since we are considering the case $\sigma
_{+}=k$, we finally get from Eq.~(\ref{a}) that
\begin{equation}
p_{r}^{\mathrm{in}}(r_{+})=-\frac{1}{8\pi r_{+}^{2}}\,,  \label{p}
\end{equation}
the desired condition. The inside pressure of a quasiblack hole with
pressure has to obey this condition. It cannot be obtained by the
straightforward limit $\varepsilon \rightarrow 0$ from the regularity
condition on the horizon of a true black hole, which as we have seen above
demands $\sigma _{+}=0$ (i.e., $p_{r}^{\mathrm{out}}(r_{+})=-\rho
^{\mathrm{
out}}(r_{+})$), see \cite{visser}. Our result represents a remarkable result
that clearly demonstrates that, although for an outside remote observer a
true black hole and a quasiblack hole are undistinguishable, in the inner
region the properties of a quasiblack hole can be very different from those
of a black hole. Our general statement that $\sigma _{+}\neq 0$ on a
quasihorizon nonextremal from inside, can be checked in the particular
examples given in \cite{bon,lzanchin1} (see also \cite{lzanchin2}) of
quasiblack holes made from presureless matter, i.e., charged dust. Indeed, for
such systems $\sigma _{+}=8\pi r_{+}\rho(r_{+})$ where $\rho(r_{+})$ is the
density of matter and its matter pressure obeys $p_r=0$ (see also
\cite{qbh}). 
Trivially, in these examples, $\rho(r_{+})$ is clearly different from
zero, so $\sigma _{+}\neq 0$, as it must. It is worth noting that the limit
discussed while checking the regularity condition can be characterized as $
\lim_{r\rightarrow r_{0}}\lim_{\varepsilon \rightarrow 0}$.

\textit{Region at the boundary, $r=r_{0}$.} We can also consider the
immediate vicinity of the boundary by taking the opposite limit: $
\lim_{\varepsilon \rightarrow 0}\lim_{r\rightarrow r_{0}}$. Then, it
follows from (\ref{uin}) that for any $\varepsilon \neq 0$ we have
that $\psi (r,r_{0})\rightarrow 0$ when $r\rightarrow r_{0}$. Thus,
$U_{\mathrm{in} }(r_{0})=V_{\mathrm{in}}(r_{0})=\varepsilon $ and the
procedure is self-consistent.

\subsubsection{Discussion: Conditions on the pressure and energy
density at the boundary and more on the regularity requirement}

\noindent \textit{(i) Conditions on the pressure and energy density at the
boundary.} We divide this discussion into two situations, when there is no
electromagnetic field and when there is one.

{(a) No electromagnetic field.} Suppose that there is no electromagnetic
field. Then, since from Eq.~(\ref{p}) the radial pressure $p_{r}$ on the
bounday is negative, we deduce that quasiblack holes with no electromagnetic
field are connected with tension on the boundary. To proceed in the
analysis, note that at a outside sphere with radius $r$, from Eq.~(\ref{m})
the mass $m(r)$ can be written as $m(r)=m(r_{+})+4\pi
\int_{r_{+}}^{r}d\bar{
r }\bar{r}^{2}\rho $. Thus, since $r_{+}=2m(r_{+})$, from Eq.~(\ref{m}) one
can write for the outside
\begin{equation}
V_{\mathrm{out}}(r)=1-\frac{r_{+}}{r}-\frac{2m_{\mathrm{out}}}{r}\text{, }
\quad m_{\mathrm{out}}=4\pi \int_{r_{+}}^{r}d\bar{r}\bar{r}^{2}\rho _{
\mathrm{out}}(\bar{r})\text{,}
\end{equation}
where the difference between a horizon and a quasihorizon has been
neglected. So, $V_{\mathrm{out}}^{\,\prime }(r)$ at $r_{+}$ is given by
\begin{equation}
V_{\mathrm{out}}^{\,\prime }(r_{+})=\frac{1}{r_{+}}\left( 1-8\pi \rho _{
\mathrm{out}}(r_{+})r_{+}^{2}\right) \,\text{.}  \label{voutprime}
\end{equation}
We recall that we are dealing with extremal quasiblack holes from
outside, since it is this kind of quasiblack holes which is free of
curvature singularities or infinite surface stresses
\cite{qbh,mass1}. Therefore. $V_{ \mathrm{out}}^{\,\prime }(r_{+})=0$,
and from Eq.~(\ref{voutprime}) we find
\begin{equation}
\rho _{\mathrm{out}}(r_{+})=\frac{1}{8\pi r_{+}^{2}}\,\text{.}
\end{equation}
From the regularity condition on the horizon of a black hole (see,
e.g., a detailed discussion in \cite{visser}) it also follows that
\begin{equation}
p_{r}^{\mathrm{out}}(r_{+})=-\rho ^{\mathrm{out}}(r_{+})\,\text{,}
\label{eqout}
\end{equation}
and so
\begin{equation}
p_{r}^{\mathrm{out}}(r_{+})=-\frac{1}{8\pi r_{+}^{2}}\,\text{.}
\end{equation}
Thus, from Eq.~(\ref{p}) one always has
\begin{equation}
p^{\mathrm{in}}(r_{+})=p^{\mathrm{out}}(r_{+})\,\text{.}  \label{inout}
\end{equation}
This means we automatically obtain a quasiblack hole with continuous
pressure on the boundary. So there is no need for a shell, certainly
an elegant result, since thin shells always imply in some type of
primary, albeit mild, discontinuity in the metric fields. On the other
hand, we are considering the case in which the matter inside is not
extremal in the sense that $V_{\mathrm{in}}^{\,\prime }(r_{+})\neq 0$
by construction. This means that a jump in density is mandatory. Jumps
in density are well handled in gravitational systems, so this means
that there is no problem. It is also important to pay attention to the
following point. In principle, quasiblack holes which are extremal
from outside, admit nonzero surface stresses and hence jumps in the
radial pressure. This conclusion was obtained in \cite{qbh,mass1} from
a general form of the metric of extremal quasiblack holes. However,
if, additionally, we take into account Einstein equations, it turns
out that for configurations which are extremal outside and nonextremal
inside, these surface stresses vanish.

{(b) Electromagnetic field.} Suppose now that there is an
electromagnetic field. Now, the pressure receives contribution from
two fields, the electromagnetic field and the matter field, so that
the radial pressure can be written as
$p_{r}=p_{r}^{\mathrm{matter}}+p_{r}^{ \mathrm{em}}$. The
electromagnetic pressure has the form $p_{r}^{\mathrm{em}
}=-\frac{q^{2}(r)}{ 8\pi r^{4}}$ where $q(r)$ is the charge enclosed
inside a sphere of radius $ r $. Bearing in mind that we are
interested in configurations which are (or tend to) extremal when
viewed from outside, we have in the limit under discussion,
$q(r_{+})=r_{+}$, in accordance with the properties of an extremal
Reissner-Nordstr\"om metric. Thus, $p_{r}^{\mathrm{em} }=-\frac{1}{
8\pi r_+^{2}}$. Then, it follows from Eq.~(\ref{p}) that
\begin{equation}
p_{r}^{\mathrm{matter}}(r_{+})=0\text{.}
\end{equation}
This situation, of existence of an electromagnetic field, is physically
preferable since it means that we can build a quasiblack hole by considering
a relativistic star with pressure obeying $p_{r}^{\mathrm{matter}}(r_{0})=0$
on the boundary and then taking the quasihorizon limit, as was done in
\cite
{jz}. In doing so, the configuration outside either represents an extremal
Reissner-Nordstr\"om quasiblack hole or tends to it as shown in \cite{qbh}.

\vskip 0.3cm \noindent \textit{(ii) More on the regularity requirement.} We
now want to emphasize the role of the regularity requirement, i.e.,
regularity in the components of the Riemann tensor and so a spacetime free
of curvature singularities. In principle, a metric in which Eq.~(\ref{f})
holds can occur without this requirement. For example, if we take $p_{r}=0$
and $\rho =\rho _{0}=\mathrm{const}$ everywhere for $r\leq r_{0}$, and
vacuum outside, an exact solution can be obtained
\cite{florides,boehmerlobo}
for which $V=1-\frac{8\pi \rho _{0}r^{2}}{3}$ and $U=\left(1-\frac{8\pi \rho
_{0}r_{0}^{2} }{3}\right)^{3/2}\left(1-\frac{8\pi \rho _{0}r^{2}}{3}
\right)^{-1/2}$. Here $r_{0}$ is the surface at which this solution matches
the outer Schwarzschild solution. One can try to obtain a quasiblack hole
from this solution by taking the limit $r_{0}\rightarrow \sqrt{\frac{3}{8\pi
\rho _{0}}}$. Then, the metric potential $U$ does indeed acquire the form
given in Eq.~(\ref{f}). However, in this limit the surface $r=r_{0}$ becomes
singular. By construction, condition (\ref{p}) is not satisfied, so the
absence of a regular quasiblack hole is justified. This, being an example in
which the outside metric is Schwarzshild rather than extremal
Reisnner-Nordstr\"om, also shows that it is much harder to find nonextremal
regular quasiblack holes than extremal ones.

\subsection{Quasiblack holes with pressure, extremal from the inside}

\subsubsection{General considerations}

In \cite{qbh} we have analyzed the properties of quasiblack holes in
which the matter in the inside region is extremal, i.e., matter for
which the energy density is equal to the charge density.  These
quasiblack holes of \cite{qbh} are thus quasiblack holes without
pressure, with extremal matter in the inside region.  Here we
generalize those results by analyzing the properties of quasiblack
holes with pressure extremal from the inside.  Extremal from the
inside means that the horizon from the inside is extremal (this is
obligatory for quasiblack holes without pressure, but not for
quasiblack holes with pressure). The horizon from the outside is
always extremal for us.

In the case we have an extremal horizon from the inside, instead of 
(\ref{uin}) we have the asymptotic form
\begin{equation}
V=\varepsilon +\kappa ^{2}(r_{0}-r)^{2}+...\text{, }  \label{vext}
\end{equation}
with $\varepsilon <<1$, and $\kappa $ being some positive quantity with
units of inverse length. See \cite{lw1,lw2} for concrete examples of this
case of quasiblack holes with nonzero pressure which represent dispersed
systems and have quasihorizons which are extremal both from inside and
outside. Note that in (\ref{vext}) we can neglect the difference between $
r_{0}$ and $r_{+}$. Inside we can distinguish two regions, the region in the
bulk of the matter $r<r_{0}$, and the region at the boundary $r=r_{0}$. We
consider now both regions separately.

\textit{Region in the bulk of the matter, $r<r_{0}$.} In this region, $
r<r_{0}$, the proper distance $l$, given by
$l=\int_{r}^{r_{0}}\frac{d\bar{r}
}{\sqrt{V}}\,$, from any point to the boundary diverges in the limit $
\varepsilon \rightarrow 0$ as it is clear from (\ref{vext}). Indeed,
defining $dl$ as the infinitesimal proper distance, one obtains in the limit
$\varepsilon \rightarrow 0$, $l\approx -\frac{1}{2\kappa }\,\ln \varepsilon
$. 
It is useful to proceed along the same lines as in Sec. III A but now,
because of the different asymptotic form of $V$, it is more convenient to
rewrite $\psi $ in another form,
\begin{equation}
U_{\mathrm{in}}=V_{\mathrm{in}}P_{1}P_{2}\,P_{3}.  \label{u3}
\end{equation}
Using the definition (\ref{sigmag}), we can rewrite the function $\psi $ in
(\ref{uin}), in this limit, as
\begin{equation}
P_{1}=\exp (2\psi _{1})\;\text{, }\quad 2\psi
_{1}=\int_{r_{0}}^{r}d\bar{r}\,
\frac{\sigma _{0}}{V_{\mathrm{in}}({\bar{r}})}\,,  \label{p1}
\end{equation}
\begin{equation}
P_{2}=\exp (2\psi _{2})\;\text{, }\quad 2\psi
_{2}=\int_{r_{0}}^{r}d\bar{r}\,
\frac{\sigma (\bar{r})-\sigma _{0}-{\bar{\sigma}}_{0}^{\,\prime }(\bar{r}
-r_{0})}{V_{\mathrm{in}}({\bar{r}})}\,\text{,}  \label{p2}
\end{equation}
\begin{equation}
P_{3}=\exp (2\psi _{3})\text{, }\quad 2\psi _{3}=\int_{r_{0}}^{r}d\bar{r}\,
\frac{\sigma _{0}^{\,\prime }(\bar{r}-r_{0})}{V_{\mathrm{in}}({\bar{r}})}
\text{ ,}  \label{p3}
\end{equation}
where again a $^{\,\prime }$ denotes a derivative with respect to the
argument. Consider each term on (\ref{u3}) separately in the limit $
\varepsilon \rightarrow 0$. In the first term, the integral is of the
order $ \varepsilon ^{-1/2}$. To make the whole expression finite, we
must conclude that $\sigma _{0}$ $\approx \sigma _{+}$ is also of the
same order to compensate these divergences, namely, $\sigma
_{+}\lesssim O(\sqrt{ \varepsilon })$, i.e., $p_{r}+\rho \lesssim
O(\sqrt{\varepsilon })$, see \cite{qbh} (Sec.II.A.d) for the analogous
result for extremal charged dust.  The second term remains finite
since near $r_{0}$ both the numerator and denominator are proportional
to $(\bar{r}-r_{0})^{2}$ in the limit under discussion. Consider now
the third term. We are discussing the region $ r<r_{0}$. Thus, if
$\sigma _{+}^{\,\prime }>0$, it is seen that in the region under
discussion $\psi _{3}\rightarrow +\infty $, $P_{3}\rightarrow +\infty
$, $U_{\mathrm{\ in}}\rightarrow +\infty $. Such a behavior has
nothing to do with a quasiblackhole and should be rejected. Therefore,
we must have $\sigma _{+}^{\,\prime }\leq 0$. Because of the
logarithmic behavior of the integral, we can represent $\psi _{3}$ in
the form
\begin{equation}
2\psi _{3}=\frac{\sigma _{+}^{\,\prime }}{2\kappa ^{2}}\left( \ln \left(
(r-r_{0})^{2}+\frac{\varepsilon }{\kappa ^{2}}\right) -\ln \left( \frac{
\varepsilon }{\kappa ^{2}}\right) \right) +2\psi _{33}
\end{equation}
where $2\psi _{33}$ is finite in the limit under discussion (cf.
Eq.~(\ref{q})). Then, we can write the metric as (cf. Eq.~(\ref{g})),
\begin{equation}
ds^{2}=-V(r)\left( (r-r_{0})^{2}+\frac{\varepsilon }{\kappa ^{2}}\right) ^{
\frac{\sigma _{+}^{\,\prime }}{2\kappa ^{2}}}\,g(r)\,dT^{2}+\frac{dr^{2}}{
V(r)}+r^{2}\left( d\theta ^{2}+\sin ^{2}\theta d\phi ^{2}\right)
\label{nm}
\end{equation}
where $T=t\left( \frac{\varepsilon }{\kappa ^{2}}\right) ^{-\frac{\sigma
_{+}^{\,\prime }}{4\kappa ^{2}}}$, and $g=\exp (2\psi _{33}+2\psi _{1}+2\psi
_{2})$ is finite. The concrete form of the metric potentials in the interior
region is model dependent, see examples in \cite{qbh} (see also \cite
{lw1,lw2} for extremal pressure systems, and \cite{lzanchin2,extdust} for
extremal pressureless systems). We can also discuss the regularity of the
Riemann tensor as we did in the nonextremal case. Using (\ref{nm}) and
(\ref{vext}) in (\ref{e}) and (\ref{ge})
gives in the limit $r\rightarrow r_{0}$ that
\begin{equation}
{Q}\approx \frac{\sigma _{+}^{2}}{4\,V}\text{.}
\end{equation}
so the only possible choice is indeed
\begin{equation}
\sigma _{+}=O(\sqrt{\varepsilon })\rightarrow 0\text{,}
\end{equation}
as already found.

\textit{Region at the boundary, $r=r_{0}$.} This region is in the
immediate vicinity of the boundary (which tends to the quasihorizon in
the limit under discussion). In this region, by definition, the proper
distance $l$ remains finite since, although the double root of V is
being approached, the limit of integration shrinks.the limit of
integration shrinks. We assume that the metric is well-defined, with
$2\psi $ being finite in the vicinity of $r_{0}$. Then, bearing in
mind that $\sigma _{+}\lesssim O(\sqrt{\varepsilon })$ as found above,
neglecting a weak dependence of $\sigma /r$ on $r$, so that $\frac{
\sigma }{r}\approx {a\kappa ^{2}}\sqrt{\varepsilon }$ for some
constant $a$, we can write near the quasihorizon $r_{+}$,
\begin{equation}
\sigma _{+}\approx {a\kappa ^{2}}\,r_{+}\sqrt{\varepsilon }\text{.}
\label{c}
\end{equation}
Here, $a\geq 0$ since, as discussed above, near the quasihorizon we want to
have $\sigma^{\,\prime}<0$ and $\sigma>0$. In the limit $\varepsilon =0$ we
obtain from (\ref{c}) the regularity condition for the quasihorizon, the
condition being $\sigma _{+}=0$, which by Eq.~(\ref{a}) means
\begin{equation}
p_{r}(r_{+})=-\rho (r_{+})\,.  \label{eqin}
\end{equation}
This regularity condition is the same as for usual, i.e. true, horizons, see
e.g. \cite{visser}. Thus, if a quasihorizon is extremal from inside, the
regularity condition (\ref{eqin})$\ $similar to that for black hole 
(\ref{eqout}) is reproduced, in contrast to the situation with the quasihorizon
nonextremal from inside. To obtain the metric in this limit we make the
substitution $r=r_{0}-\sqrt{\frac{\varepsilon }{\kappa ^{2}}}\,y$. Then the
metric is
\begin{equation}
ds^{2}=-(1+y^{2})\mathrm{e}^{-(a\kappa \,r_{+}\arctan y)}\,dT^{2}+\frac{1}{
\kappa ^{2}}\frac{1}{1+y^{2}}\,dy^{2}+r_{0}^{2}\left( d\theta ^{2}+\sin
^{2}\theta d\phi ^{2}\right) \,.  \label{br}
\end{equation}
We have used $t=\frac{T}{\sqrt{\varepsilon }}$ to absorb the factor $
\varepsilon $ into $g_{00}$, a procedure that is typical of quasiblack holes
\cite{qbh}. The metric (\ref{br}) is a slight generalization of the
Bertotti-Robinson metric. To see it, we note that for a pure electromagnetic
situation one has $\kappa ^{2}=\frac{1}{r_{+}^{2}}$ and $a=0$, so that
(\ref{br}) 
coincides with the Bertotti-Robinson metric. The proper distance $l=
\frac{1}{\kappa }\int_{0}^{y}\frac{d\bar{y}}{\sqrt{1+y^{2}}}$ is finite for
any $y$ but it diverges in the limit $y\rightarrow \infty $, so we obtain an
infinitely long tube.

\subsubsection{Discussion: conditions on the pressure and energy density at
the boundary \newline
and a proposition}

\noindent \textit{(i) Conditions on the pressure and energy density at the
boundary.} Since the value of $\rho (r_{+})$ is fixed by the condition $
V^{\,\prime }(r_{+})=0$ both from outside and inside, both radial pressure
and density are continuous, in contrast to the nonextremal case from inside
where the density is discontinuous. In the situation (a) there is no
electromagnetic field then the quasihorizon is supported by matter tension,
in the situation (b) there is an electromagnetic field the matter pressure
is equal to zero at the quasihorizon, both results can be deduced as before.

\vskip0.3cm \noindent \textit{(ii) A proposition.} From the above
considerations an interesting result follows. In order to have a
well-defined $U$, and thus a well-defined metric, we need to have $\sigma$
defined in Eq.~(\ref{sigmag}) obeying ${\sigma }>0$ in some vicinity of the
quasihorizon. Since on the quasihorizon itself $\sigma=\sigma _{0}=\sigma_+
\rightarrow 0$, we must have $\sigma _{0}^{\,\prime }\leq 0$ as is explained
above. But from (\ref{sigmag}), $\sigma =8\pi r\bigl(p_{r}(r)+\rho
(r)\bigr)$. 
Thus, we can state the following proposition: (i) One cannot build an
extremal quasiblack hole entirely from phantom matter, i.e., matter with the
null energy condition violated everywhere inside, $p_{r}+\rho <0$. (ii) In
case there is phantom matter, it cannot border the quasihorizon but must lie
inside the inner region only. Thus, at least in some vicinity of the
quasihorizon the null energy condition is satisfied everywhere, so that $
p_{r}+\rho \geq 0$.

For a discussion of the energy conditions within the related context of
regular black holes see \cite{zasl}. Alternation of regions with normal and
phantom matter is discussed in \cite{bs} in another context.

\section{Conclusion}

\label{conc}

We have studied extremal quasiblack holes, as seen from the outside, with
nonzero pressure and have shown how these objects are attainable on general
grounds. From the inside these quasiblack holes can have nonextremal and
extremal quasihorizons. The total pressure at the matter boundary is less or
equal to zero and it is always continuous there. In the situation 
where there is
an electric field the matter pressure is zero at that boundary. The density
behaves as expected, either showing a jump at the boundary in the
nonextremal case or being continuous in the extremal case. The regularity
conditions for the nonextremal inside case is completely different from the
regularity condition for the usual regular black holes, whereas the
regularity conditions for the extremal inside case can be obtained from the
known regularity conditions for the usual regular black holes. For the
extremal inside case we show that the quasiblack holes cannot be made from
phantom matter at the quasihorizon. Further properties that one can envisage
depend on the particular model under study, see \cite{jz,felice1,felice2}
for the nonextremal inside case and \cite{lw1,lw2} for the extremal inside
case. In our previous studies \cite{qbh,wh,mass1,mass2,entr,extrentr} we
have shown that quasiblack holes with nonextremal and extremal
quasihorizons for the outside are distinct entities and must be considered
as such when one studies them. Here we have shown that the same holds for
quasiblack holes with nonextremal and extremal quasihorizons for the
inside. They have to be carefully considered as separate entities with
distinct properties.

\begin{acknowledgments}
This work was partially funded by Funda\c c\~ao para a Ci\^encia e
Tecnologia (FCT) - Portugal, through project Nos. PTDC/FIS/098962/2008 and
CERN/FP/109276/2009.
\end{acknowledgments}

\end{document}